\documentclass{aa}
\usepackage{graphicx}
\usepackage{amssymb}

\usepackage{color}
\usepackage{natbib}
\usepackage{tabularx}

\newcommand{\HESSJ}{HESS~J1507-622}

\def\arcsec{\hbox{$^{\prime\prime}$}}

\begin{document}

\title{The environment of the unidentified gamma-ray source \HESSJ}

\author{Wilfried F. Domainko}

\institute{Max-Planck-Institut f\"ur Kernphysik, Saupfercheckweg 1, D-69117 Heidelberg, Germany}

\offprints{\email{wilfried.domainko@mpi-hd.mpg.de}}

\date{}
 
\abstract 
{The nature of the gamma-ray source \HESSJ\ that is located significantly
off-set from the Galactic plane is not ascertained to date.}
{Identifying the environment of an enigmatic object may help to constrain
its nature.}
{The path of the line of sight of \HESSJ\ through the Galaxy is compared
to the characteristic length scales of stellar populations of different
ages. Furthermore, for this object, the energy density in particles is contrasted to the magnetic field
energy density and constraints on the distance based on equipartition between
these two components are calculated.}
{The line of sight of \HESSJ\ reaches a minimum distance
to the Galactic center at around a galactocentric distance of 5.3~kpc 
at about 300~pc off the Galactic disc. 
This location coincides with the scale length
and width of stars with an age of 1.2~Gyr which could in principle
be an indication that \HESSJ\ is connected to a stellar population of similar age. 
For such a case the source appears
to be strongly particle dominated. In a leptonic scenario, if a magnetic 
field in the source
of 1~$\mu$G is assumed, equipartition between magnetic field and particles
would be realized at a distance of $\gtrsim1$~Mpc. This could indicate
an extragalactic origin of this object. However, an extragalactic origin 
is challenged by the extension of the source.}
{The environment of \HESSJ\ still remains elusive.
For the case where this source belongs to a new class of gamma-ray emitters, 
the distribution of related objects (if existing)
may help to settle the respective environment and distance scale.}


\authorrunning{W. Domainko}

\titlerunning{The environment of \HESSJ}

\maketitle


\section{Introduction}

Several unidentified very-high energy (VHE, E$>$~100~GeV) gamma-ray sources have been found by H.E.S.S. in a
survey of the inner galaxy \citep{aharonian2008}. One particular source appears to be exceptional
since it is the only unidentified sources that is located significantly off-set from the Galactic
plane \citep[3.5$^{\circ},$][]{acero2011}. This VHE emitter is connected to a high energy 
(HE, 100~MeV~$<$~E~$<$~100~GeV) counterpart \citep[based on 24 month of \emph{Fermi}-Lat data,][]{nolan2012} 
and potentially also to a faint, diffuse X-ray
counterpart \citep{acero2011}. The nature of the source has been disputed since its 
discovery. \citet{domainko2011a} concluded that the compactness of the source disfavors a
supernova remnant (SNR) scenario. \citet{domainko2012} analyzed a larger
HE data set of 34 month from the \emph{Fermi} satellite and found that the gamma-ray
spectral energy distribution (SED) appears to be rather flat from the GeV to the TeV regime.
In this paper it was further discussed that the measured gamma-ray SED, the compactness of the source 
and its off-set location from the Galactic plane challenge a pulsar wind nebula (PWN) scenario
for HESS~J1507-622. \citet{vorster2013} fitted a PWN model to the multi-wavelength SED 
\citep[using the result of 24 month of Fermi exposure obtained by][]{nolan2012} and 
concluded in favor for a PWN scenario. \citet{acero2013} finally 
analyzed 45 month of Fermi data above 10 GeV and placed
this object in their non-PWN sample. 

In this paper an alternative approach is adopted to explore the nature of
HESS~J1507-622. Its exceptional position is used to investigate
the properties of the stellar environment along the line of sight of the
source. Similar approaches have been applied to constrain the origin of enigmatic objects in the past.
Studies of the stellar ages and stellar environments have extensively been applied
to investigate the nature of the progenitors of supernovae~Ia \citep[e.g.][]{maoz2011} and
short gamma-ray bursts \citep[e.g.][]{berger2014}. Prospects for the origin of the 
VHE gamma-ray source HESS~J1747-248 have been presented based on its location in the vicinity
of the Galactic globular cluster Terzan~5 \citep{abramowski2011,domainko2011b}.
Finally, host galaxies of extragalactic gamma-ray emitters of BL~Lac-type have
been studies in depth \citep[e.g.][]{cheung2003,shaw2013}.

This paper is organized in the following way: In Sec.~\ref{sec:los}
the Galactic stellar population along the line of sight of \HESSJ\ is
explored. In Sec. \ref{sec:obj} other potential sources of VHE emission
at Galactic off-plane locations and their relation to \HESSJ\ are discussed.
In Sec. \ref{sec:equi} constraints on the distance to \HESSJ\ based on
equipartition between the energy densities in particles and magnetic fields
are evaluated. And in Sec. \ref{sec:ext} prospects for an extragalactic scenario
for \HESSJ\ are explored.

\section{Parent stellar populations along the line of sight of HESS~J1507-622}
\label{sec:los}

\begin{figure*}[ht]

\includegraphics[height=6.6cm]{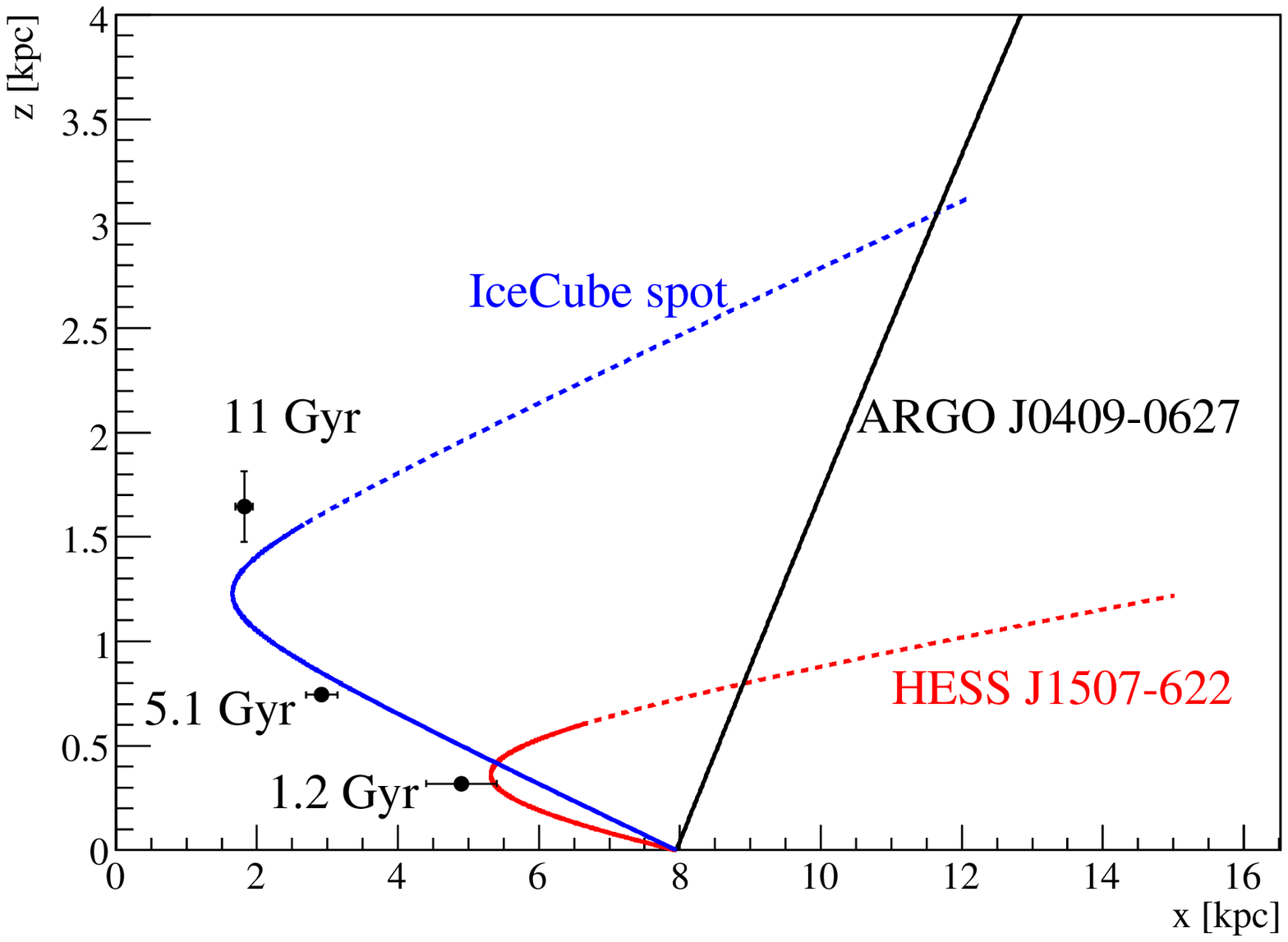}
\includegraphics[height=6.6cm]{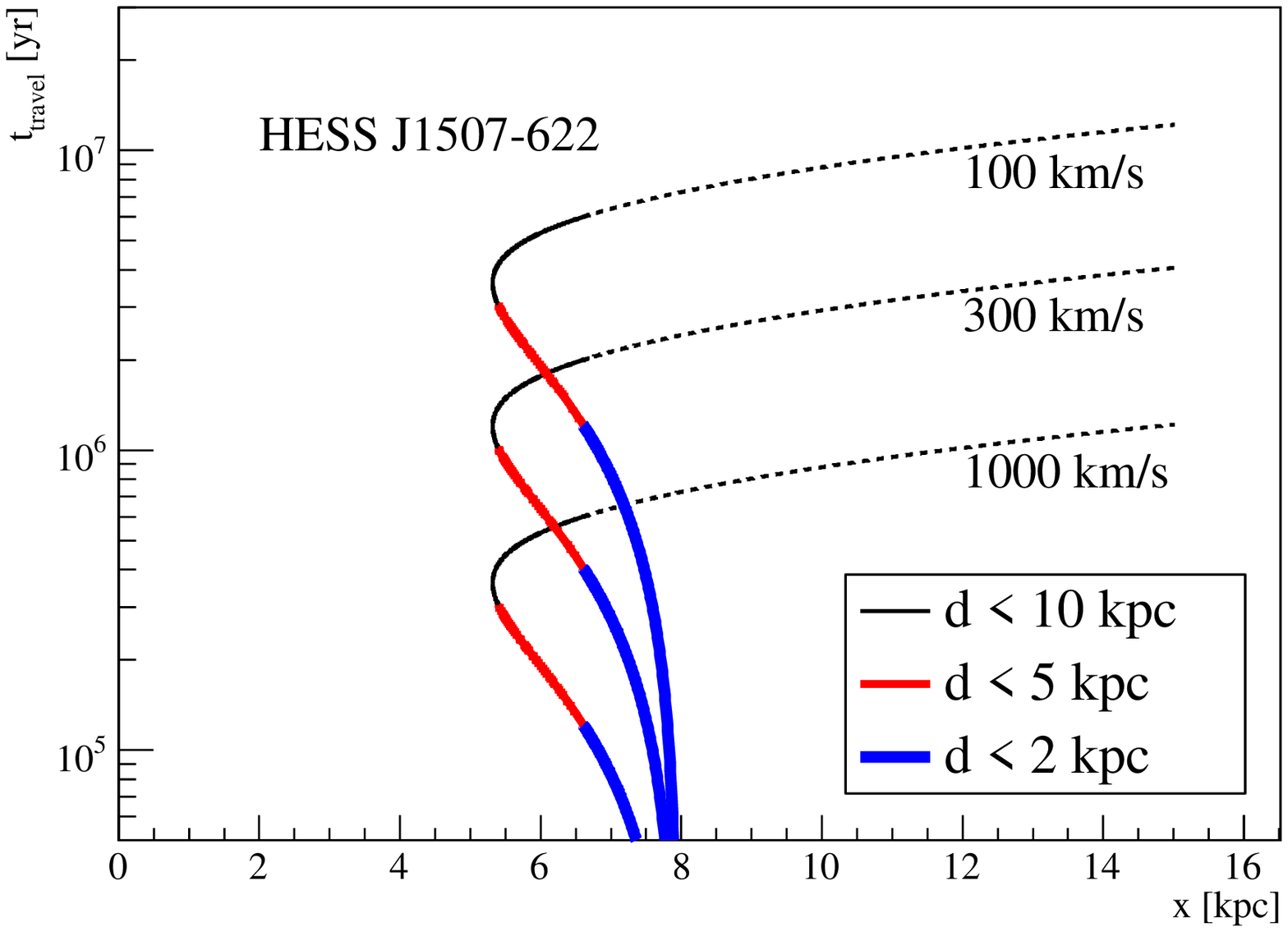}
\caption{The path of the line of sight of \HESSJ\ through the Galaxy is shown in this figure.
Left panel: Here the path of the line of sight is compared to the scale length of
stellar populations with various ages ($x$ is the galactocentric radius in the Galactic plane
and $z$ is the absolute height above the plane). For \HESSJ\ the region where the path of the
line of sight is closest to the Galactic center coincides with the scale length of
a stellar population with age of about 1.2~Gyr \citep{stinson2013}. For comparison also the line of sights
for ARGO~J0409-0627 and the spot with highest probability for being a VHE neutrino
source are also shown. Solid lines indicate a source distance of less than 10~kpc
(the distance to ARGO~J0409-0627 within an absolute height above the Galactic plane
of $\lesssim$~4kpc is always shorter than 10~kpc).
Right panel: Times of travel for a high space-velocity stellar population with various
velocities perpendicular to the Galactic plane are shown
along the line of sight of HESS~J1507-622. The blue thick solid line indicates a source 
distance of less than 2~kpc, the red medium thick solid line indicates a source distance of 
less than 5~kpc and the black thin solid line indicates a source distance of less than 10~kpc.}
\label{figure:location}
\end{figure*}

\subsection{Parent stellar population}

In the Galaxy, old stars lie in thick distributions around the plane with short
disk scale length \citep[e.g.][]{stinson2013}. For younger stellar populations
the width of the distribution decreases and the disk scale length increases.
The compactness of \HESSJ\ suggests a distance to the object of several kpc
\citep[e.g.][]{hinton2009,domainko2012} that is comparable to the characteristic
length-scales of the stellar population of the Galaxy.
Therefore, the line of sight through the Galaxy of \HESSJ\ can in principle be compared to the
characteristic length scales of the distributions of stellar populations
of different ages (see Fig. \ref{figure:location}, left panel). This line of sight
reaches a minimal distance to the galactic center at a galactocentric
distance of 5.3 kpc 
at about 300~pc off the Galactic disc 
\citep[assuming a Galactic center -- sun distance of 7.94 kpc,][]{eisenhauer2003}. 
For a bulge-like stellar distribution, there the line of sight passes through
the region with the highest stellar density.
The location of minimal galactocentric distance coincides with the scale length
and width of stars with an age of 1.2 Gyr \citep{stinson2013} which could in principle
be an indication that \HESSJ\ is connected to a stellar population of similar age. For comparison,
for supernovae~Ia, a prompt channel is found (age $<$ 420~Myr) together with a
population that is delayed by $>$~2.4~Gyr \citep{maoz2011}.
For a further comparison, the stellar ages connected to short gamma-ray bursts span from a few tens of Myr to 
about 4~Gyr \citep{berger2014}. It has to be noted that a few off-plane objects
of the same type (if existing) would be needed for a more conclusive comparison 
with stellar populations of different ages (see Sec. \ref{sec:obj}).

\subsection{High space-velocity stellar population}

In the previous paragraph it has been assumed that \HESSJ\ shares the typical velocity
distribution of its parent stellar population. However, some particular objects travel
with much larger space velocities \citep[pulsars for example show a mean three dimensional velocity
of $380^{+40}_{-60}$~km~s$^{-1}$,][]{faucher2006}. For such a case constraints on the 
age of the VHE emitter can be estimated from the time of travel to the observed location
if it is assumed that the object was born close to the Galactic plane. The time of travel
as a function of location in the Galaxy for \HESSJ\ is shown in Fig. \ref{figure:location}
(right panel). Estimated travel times range from $\gtrsim 10^5$~years for velocities perpendicular
to the Galactic plane of $\approx 1000$~km~s$^{-1}$ to $\approx 10^7$~years for moderate
velocities of $\approx 100$~km~s$^{-1}$ and multi-kpc distances. Only the shorter estimates
of the travel times would be compatible with the ages of observed PWNe \citep{kargaltsev2013}.
However, massive high-velocity stars with lifetimes $\gtrsim 10^7$~years would be able to reach such 
off-plane locations. For constraints on the distance to \HESSJ\ see Sec. \ref{sec:equi}.

\section{Other potential sources of VHE emission at Galactic off-plane locations}
\label{sec:obj}

In contrast to other unidentified VHE gamma-ray sources that are located at the Galactic plane
\citep[see][]{aharonian2008}, \HESSJ\ is located significantly off-set from it.
This object could simply be an outlier to the usual distribution of sources, or
it could be a representative of a class of VHE gamma-ray emitters with broader
distribution around the Galactic plane. In this section indications for an
off-plane source population are explored. Furthermore, properties of potential VHE emitters at locations
off the Galactic plane are briefly reviewed. 
The relation between these source candidates
and \HESSJ\ is unclear at the moment. However, if some of these objects are
of similar type as HESS~J1507-622, their position on the sky may help to
further constrain the environment of this source.

\subsection{ARGO J0409-0627}

Recently the ARGO-YBJ collaboration has reported on a survey of the northern sky in the 
VHE gamma-ray regime \citep{bartoli2013}. In this survey one source candidate without clear identification
has been found with large off-set from the galactic plane: ARGO~J0409-0627. It has to be noted 
that this source candidate has the lowest post-trial significance ($< 3 \sigma$) of all presented sources.
At a position of l~=~198.5$^\circ$, b~=~-38.7$^\circ$ it seems not to follow the distribution of
Galactic stellar populations (see Fig. \ref{figure:location} left panel). Its position coincides with a group
of galaxies \citep[SDSS~J040922.92-062636.4, SDSS~J040921.13-062936.5, SDSS~J040912.63-062546.6,
SDSS~J040914.54-063030.7,][at a common redshift of 0.12]{adelman2008}\footnote{http://ned.ipac.caltech.edu/},
however a physical association is unclear at the moment.

\subsection{IceCube neutrino sources}

The IceCube collaboration has detected evidence of extraterrestrial neutrinos in the TeV-PeV regime \citep{aartsen2013}.
With the current data, the distribution of these neutrino events is consistent with an isotropic distribution on the
sky, with a spot of highest probability for being a neutrino source located at  l~=+12$^\circ$, b~=~-9$^\circ$.
In the case that the VHE neutrino sky comprises sources rather than truly diffuse emission it is expected that neutrino 
sources are linked to VHE gamma-ray sources \citep[e.g.][]{kappes2007} and
for the detected neutrino events a correlation with unidentified VHE gamma-ray sources has already been suggested \citep{fox2013}.
For the purpose of this paper it is assumed that the  VHE neutrinos originate from discrete sources.
Most relevant for the discussion in this paper is the fact that the distribution
of extraterrestrial neutrinos seems not to be strongly peaked at the galactic plane. 
Additionally, the gamma-ray SED of \HESSJ\ can reasonably be fit by a hadronic model \citep{domainko2012}. 
If this is the actual emission process 
of the source, that would imply neutrino emission from objects of similar type.
However, so far there is no positional coincidence of any neutrino event 
with the location of HESS~J1507-622. 

In Fig. \ref{figure:location} (left panel) the line of sight of the neutrino spot
through the galaxy is compared to the distribution of stellar populations of various ages. If located close
to its minimal distance to the Galactic center the position of this spot would coincide with the characteristic length
scale of a very old stellar population. However is has to be noted that the positional uncertainties of
shower-like neutrino events is rather large \citep[$\gtrsim 10^\circ$,][]{aartsen2013}. 
It has to be further noted that the distance scale to the VHE neutrino emission sites is not known to date.
For implications of an
extragalactic scenario see Sec. \ref{sec:ext}.

\section{Comparison between the energy density in particles and the magnetic field}
\label{sec:equi}

\begin{figure*}[ht]
\includegraphics[height=6.6cm]{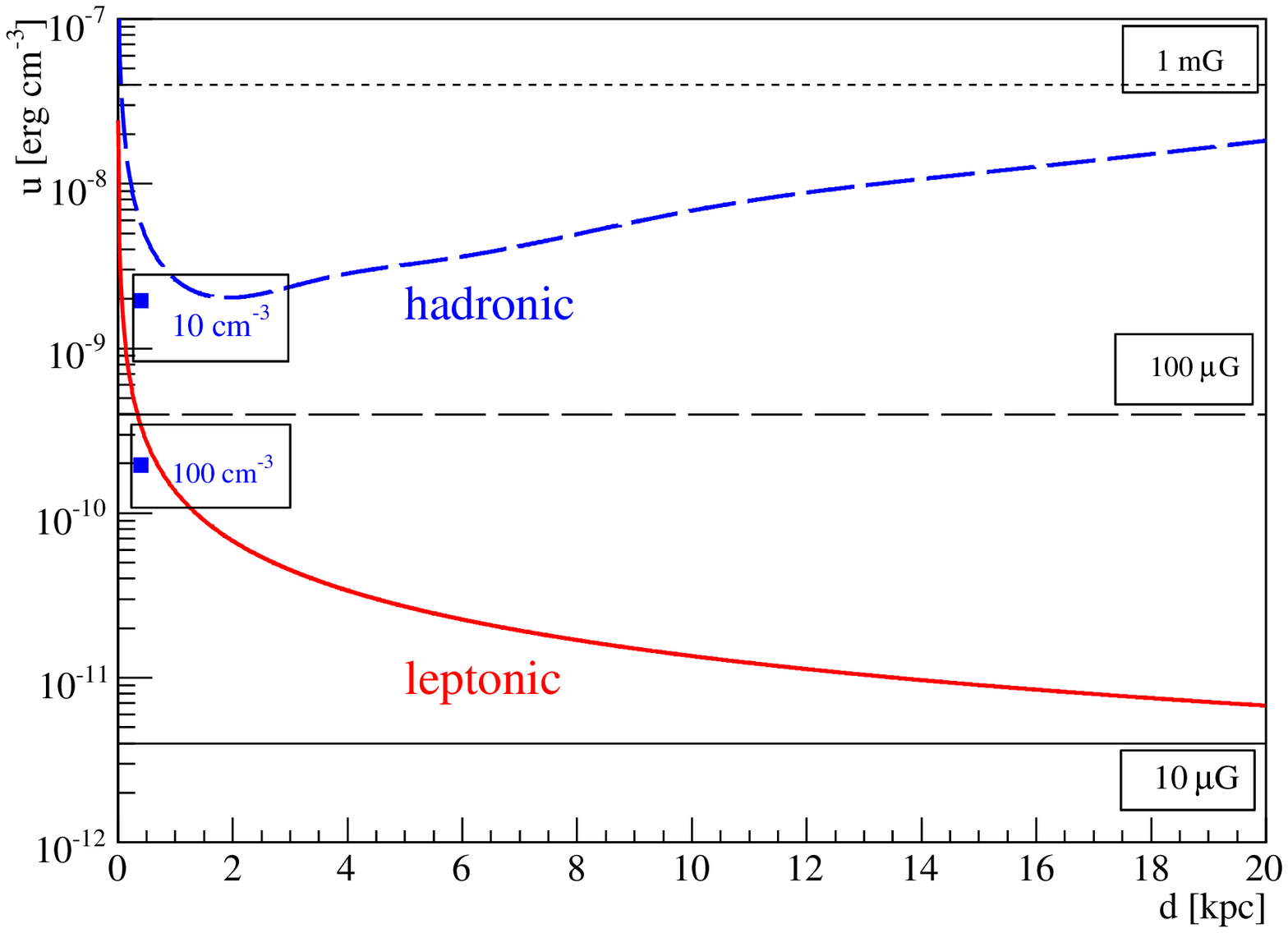}
\includegraphics[height=6.6cm]{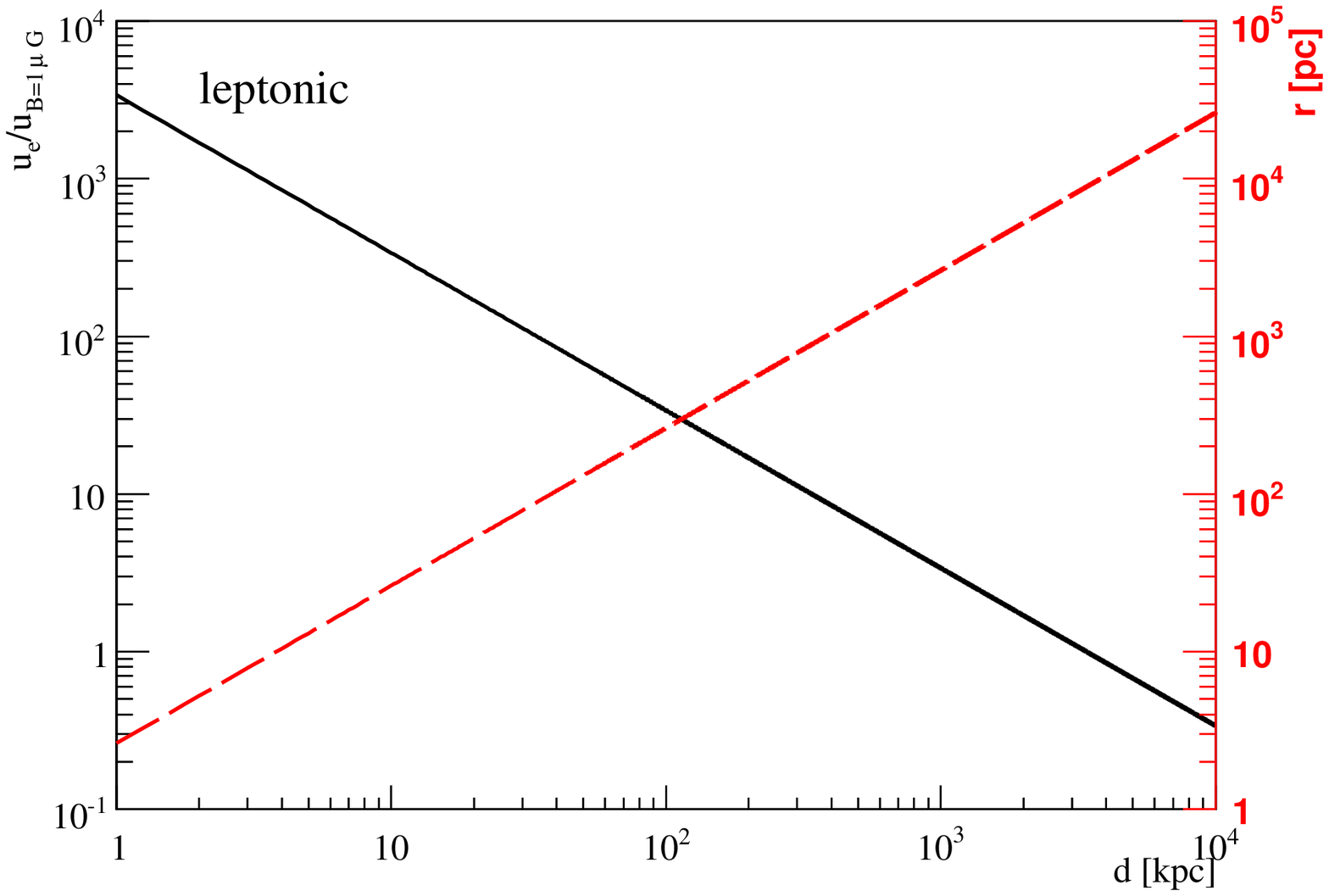}
\caption{Left panel: the energy density in particles in \HESSJ\ as a function of distance
is shown (for details see main text). Additionally, the energy density for a hadronic scenario
for the case where the source is located on the edge of a molecular cloud at a distance of 400~pc
\citep{acero2011} for a density of target material of 10~cm$^{-3}$ and 100~cm$^{-3}$ is plotted
as blue dots. For comparison, the energy density of magnetic fields with strength 10~$\mu$G,
100~$\mu$G and 1~mG is given. Right panel: The ratio of energy density in particles and
energy density in the magnetic field as a function of distance is shown as a solid black line for the case of 
1~$\mu$G. 
These two energy densities roughly equal at a distance of about 4~Mpc. On the right hand y-axis
the source extension as a function of distance is shown with a dashed red line. 
At a distance of 4~Mpc, the source would extend to $\approx10$~kpc.}
\label{figure:equi}
\end{figure*}

In this section the particle energy density of \HESSJ\ as function of distance is derived and is compared
to the energy density in the magnetic field. Furthermore, constraints on the distance of the object
are evaluated based on the assumption that the particle energy density approximately equals the energy
density in the magnetic field. Evolved sources may approach this kind of equilibrium state
\citep[see][for a discussion]{longair1994}.

\subsection{Leptonic scenario}

In a leptonic scenario, the energy in electrons energetic enough to produce gamma-ray emission through inverse-Compton
up-scattering of cosmic microwave background photons amounts to $E_\mathrm{e}(d) = 3 \times
10^{47} (d/1\,\mathrm{kpc})^2$\,erg with $d$ being the distance to the
source \citep[if a considerable fraction of the total energy in particles is stored in lower energy
electrons, the total energy increases accordingly, see][]{domainko2012}.  
For comparison, \citet{vorster2013} found an energy in particles of about $1.5 \times 10^{48}$~erg
for a distance of 6~kpc \citep[about a factor of 7 smaller than the estimate of][]{domainko2012}.
Since \citet{domainko2012} also included spectral points below 10~GeV for the estimate of the energy
in electrons, this value is used for the purpose of this paper.
The volume of the object
scales with $V(d) = 4 \pi /3 \times (d \times \mathrm{tan}\, \theta)^3$ with $\theta = 0.15^\circ$ being the angular radius
of the source. Using these two relations, the energy density in electrons is given by 
$u_\mathrm{e}(d) = E_\mathrm{e}(d)/V(d)$. The particle energy density scales with distance as
$d^{-1}$. It is plotted in Fig. \ref{figure:equi} left panel and there it is compared to
the energy density in magnetic fields of strength 10~$\mu$G,
100~$\mu$G and 1~mG with $u_\mathrm{B} = B^2 /8 \pi$. It is found that for a Galactic origin of \HESSJ\ ($d \lesssim 20$~kpc)
the particle energy density would exceed the energy density in a magnetic field of 10~$\mu$G. 

\subsection{Hadronic scenario}

For a hadronic scenario the energy in protons is given by $E_\mathrm{p}(d) = 4 \pi d^2 \times F_\gamma \times \tau_\mathrm{pp}(d)$.
Here $F_\gamma \approx 8.2 \times 10^{-12}$\,erg\,cm$^{-2}$\,s$^{-1}$ is the gamma-ray flux above 1~GeV \citep{domainko2012}
and, $\tau_\mathrm{pp} = 3 \times 10^{7} n^{-1}(d)$~years, is the cooling time of hadronic cosmic rays for a given density
of target material $n$. The density of target material varies with distance to HESS~J1507-622. 
Since the line of sight towards \HESSJ\ moves to larger Galactic heights below the plane 
with increasing distance $d$, the density 
of target material as a function of distance drops monotonically \citep[see][]{domainko2011a}.
Here the density of target material as a function of absolute height off the Galactic plane $z$ is adopted to be \citep{dickey1990}:

\begin{equation}
n(z) = 0.395\, e^{\frac{-z^2}{127 \mathrm{pc}}} + 0.107\, e^{\frac{-z^2}{318 \mathrm{pc}}} + 0.064\, e^{\frac{-z}{403 \mathrm{pc}}}\, \mathrm{cm^{-3}}
\end{equation}

If now the relation $z = d \times \mathrm{sin}\, 3.5^\circ$
($3.5^\circ$ being the angular distance of \HESSJ\ from the Galactic plane) is substituted in $n(z)$,
the density of target material as function of distance and thus also $E_\mathrm{p}(d)$ can be calculated.
By using the volume of the source $V(d)$ from the previous paragraph,
the energy density in particles is given by $u_\mathrm{p}(d) = E_\mathrm{p}(d)/V(d)$.
It is plotted in Fig. \ref{figure:equi} left panel. Additionally, since \HESSJ\ is located at the edge of a molecular cloud
at a distance of 400~pc \citep[see][]{acero2011}, the energy density for this distance assuming a density of target material
of 10~cm$^{-3}$ and 100~cm$^{-3}$ is also shown in Fig. \ref{figure:equi}. From Fig. \ref{figure:equi} left panel it is evident that 
for a hadronic scenario the particle energy density for a Galactic origin of \HESSJ\ would in most cases
exceed the energy density in a magnetic field of 100~$\mu$G.

\subsection{Constraints on the distance}

The calculated energy density in particles can be compared to the energy density in the magnetic field
estimated for HESS~J1507-622.
From SED fitting \citet{domainko2012} and \citet{vorster2013} constrained the magnetic field for a leptonic scenario in \HESSJ\
to $B \lesssim 1~\mu$G. The resulting magnetic field energy density $u_\mathrm{B} \lesssim 4 \times 10^{-14}$~erg~cm$^{-3}$ 
is several orders of magnitude lower than the energy in particles for typical Galactic distances (see Fig. \ref{figure:equi} right panel). 
Thus for a Galactic origin of \HESSJ\ the source appears
to be strongly particle dominated \citep[it has to be noted that young PWNs are also particle dominated,][]{torres2014}. 
For a leptonic scenario a situation where particle energy density approximately equals the energy
density in the magnetic field is only realized for distances of $d \gtrsim 1$~Mpc \citep[even more distant if a considerable fraction of 
total energy is stored in low energy electrons that do not radiate in gamma-rays,][]{domainko2012}.
For comparison, using the model of \citet{vorster2013} an equipartition distance of 180~kpc would be found
(for $B = 1.7\, \mu$G).
Since this distance is larger than the size of the Galaxy, 
in the next paragraph extragalactic scenarios are explored for HESS~J1507-622. In a hadronic
scenario equipartition between particles and magnetic fields is not achieved
out to cosmological distances ($\gtrsim 100$~Mpc) for ambient densities of target material 
$\lesssim 0.1$~cm$^{-3}$.

\section{Prospects for an extra-galactic scenario}
\label{sec:ext}

In general extragalactic scenarios are challenging due to the large
energetics needed to power the source and due to the very extended
nature of the emission. About $3 \times 10^{53} (d/1\,\mathrm{Mpc})^2$\,erg in electrons
would be needed in a leptonic scenario and the source would spread over 
$2.6\, (d/1\,\mathrm{Mpc})$\,kpc (see Fig. \ref{figure:equi}, right panel).

\subsection{Constraints from the source extension}
\label{sec:spread}

To date only very few sources with multi-kpc extension have been detected
in the GeV band. Very extended gamma-ray sources are the Centaurus~A lobes \citep{abdo2010} 
and the Fermi bubbles \citep{su2010}. However, no
object with such a large spatial dimension has been found in the TeV regime so far. 
The extension of leptonic TeV sources
is expected to be limited to few tens of parsecs by radiation losses \citep{hinton2009}.

An alternative model for very extended sources at extragalactic distances
is provided by a scenario where VHE gamma-rays produce a halo of e$^+$ and e$^-$ pairs around 
the source \citep{aharonian1994}. More specifically, gamma-rays with energies 
$\gtrsim 100$~TeV may produce
rather compact halos of size $\lesssim 1$~Mpc.
The detection of multiple-10s of TeV to PeV neutrinos \citep{aartsen2013} indicates the
emission of gamma-rays (if not absorbed inside the source) with such energies. These sources 
(especially if they emit isotropically) could in principle
be surrounded by compact pair halos radiating in the TeV regime.

Such pair halos at a distance of $\lesssim 300$~Mpc would appear with roughly comparable
extension as HESS~J1507-622.
In a scenario where energy in particles roughly equals the energy in the magnetic field
of 1~$\mu$G (see Sec. \ref{sec:equi}), \HESSJ\ could be located at a coarsely compatible
distance if a major part of its particle energy is stored in electrons not energetic
enough to radiate in the gamma-ray regime \citep[see][]{domainko2012}. Alternatively, \HESSJ\ could
also be located at a coarsely compatible distance, if a lower
magnetic field for this source is considered for the evaluation of equipartition between
particles and magnetic field 
\citep[i.e. the diffuse X-ray source in the magenta circle of Fig. 3 of][is not the synchrotron counterpart to HESS~J1507-622]{acero2011}.

The caveat for this 
interpretation is that the nature and distance scale of the VHE neutrino emitters are not known to date
and that plasma effects my suppress the electromagnetic cascade that form pair halos 
\citep{boderick2012,schlickeiser2012}.
More generally, finally it has to be noted, that there is no evidence that equipartition is actually established
in HESS~J1507-622. 

\subsection{Constraints on the energetics}

Super-massive black holes (SMBHs) are the most common class of extragalactic VHE gamma-ray emitters
\citep[see][for a review]{hinton2009}. In this section it is evaluated whether any SMBH
located in the line of sight of \HESSJ\ would be able to power this VHE gamma-ray source.

\subsubsection{Black hole masses}

The mass of a SMBH is linked to the mass and thus luminosity of its host galaxy
\citep[e.g.][]{magorrian1998,haering2004}. Consequently from the brightness of any galaxy in the
direction of \HESSJ\ the mass of its SMBH as a function of luminosity distance $d_\mathrm{L}$ can be estimated.
Here the K-band brightness of potential galaxies is used since \HESSJ\ is located rather close to the
Galactic plane and infrared luminosities are less affected by absorption due to inter-stellar dust.
The infrared source closest to the centroid of \HESSJ\ is 2MASS~J15075554-6222336 
\citep[$m_\mathrm{K}$~=~14.39~mag, corrected for galactic extinction of $A_\mathrm{K} = 0.505$,][]{schlafly2011} 
located at an angular distance
of 230$\arcsec$. The brightest infrared source inside the extend of the VHE source is 2MASS~J15070879-6216441 
($m_\mathrm{K}$~=~11.16~mag, again corrected for extinction) located
at an angular distance of 517$\arcsec$ \citep{cutri2003}\footnote{http://simbad.u-strasbg.fr/simbad/}.
For the rest of this section it is assumed that these two sources represent galaxies harboring SMBHs.
The mass of the SMBH of these hypothetical galaxies as function of luminosity
distance can be estimated from the following relations \citep{marconi2003}:

\begin{equation}
\mathrm{log}\, M_\mathrm{BH}(d_\mathrm{L}) = 8.21 + 1.13\,(\mathrm{log}\, L_\mathrm{K}(d_\mathrm{L}) -10.9)
\end{equation}

with 

\begin{equation}
\mathrm{log}\,(L_\mathrm{K}(d_\mathrm{L})/L_\mathrm{K,\odot}) = 0.4\,(3.28 - M_\mathrm{K}(d_\mathrm{L}))
\end{equation} 

and

\begin{equation}
M_\mathrm{K}(d_\mathrm{L}) = m_\mathrm{K} + 5\,\mathrm{log}\,(d_\mathrm{L}) -5
\end{equation}

Here, $L_\mathrm{K}(d_\mathrm{L})$ is the luminosity of the host galaxy in the K-band and $M_\mathrm{K}(d_\mathrm{L})$
is the absolute K-band magnitude of the host galaxy as function of the luminosity distance
($M_\mathrm{K,\odot} = 3.28$~mag is the absolute K-band magnitude of the sun).
The masses of these hypothetical SMBHs as function of distance are plotted in Fig. \ref{figure:ext}.

\subsubsection{Comparison to the Eddington luminosity}

\begin{figure}[ht]
\centering
\includegraphics[height=6.6cm]{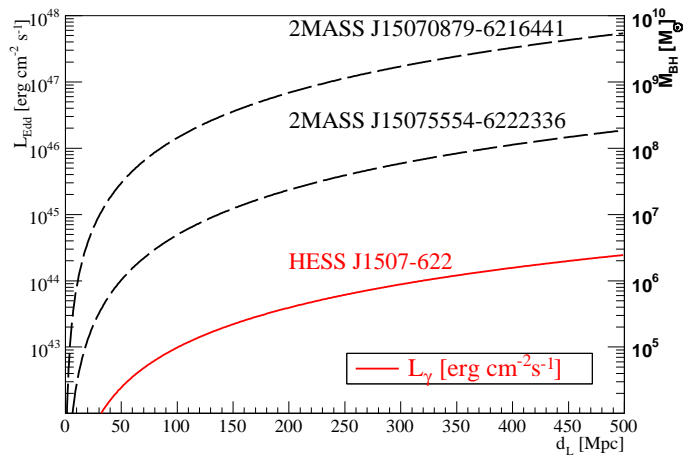}
\caption{The gamma-ray luminosity of \HESSJ\ is plotted as a function of luminosity distance
with a red solid line. For comparison, the mass and Eddington luminosity of a hypothetical SMBH in 2MASS~J15075554-6222336 and
2MASS~J15070879-6216441 is shown as a function of distance with dashed black lines
(for more details see main text).}
\label{figure:ext}
\end{figure}

With these calculated SMBH masses, constraints on their luminosities in electromagnetic
radiation can be made.  
The maximum possible continuous luminosity radiated by
a black hole is given by the Eddington luminosity that scales with black hole mass $\mathrm{M_\mathrm{BH}}$ as 
$L_\mathrm{Edd} \approx 10^{38} (\mathrm{M_\mathrm{BH}}/1 \mathrm{M_\odot})$~erg~s$^{-1}$. 
With this input $L_\mathrm{Edd}$ as a function of luminosity distance for
any potential SMBH in the direction of \HESSJ\ can be calculated and can be
compared to the gamma-ray luminosity of the source (see Fig. \ref{figure:ext}).
Since \HESSJ\ is an extended object it is expected that the emission is not beamed
and can thus be compared to $L_\mathrm{Edd}$.
It is found that \HESSJ\ would have to radiate on the level of 1\% (2MASS~J15075554-6222336)
and 0.05\% (2MASS~J15070879-6216441) of $L_\mathrm{Edd}$ of a SMBH in these
hypothetical galaxies. To conclude, a scenario where \HESSJ\ is powered by a SMBH
is in principle energetically possible. However, in this case, the extension of the source
still has to be explained (see Sec. \ref{sec:spread}). Nevertheless, if similar sources
as \HESSJ\ will be found at large Galactic latitudes, this may indicate
an extragalactic scenario for these objects (see Sec. \ref{sec:obj}).

\section{Summary and outlook}

In this paper the environment of the unidentified off-plane gamma-ray source \HESSJ\ is
explored. Since the location on the sky of this object is unique with respect
to the positions of other unidentified VHE gamma-ray sources, examining
the environment may give additional insights in its nature.
However, with the currently available informations the surroundings of
\HESSJ\ could not be identified.

For a Galactic origin of this source its location may indicate 
a parent stellar population as old as 1~Gyr. In this case the source appears to be
rather strongly particle dominated. In principle, particle dominance seems to
be possible for
Galactic VHE gamma-ray emitters, with young PWNe also showing this feature.
However, in a leptonic scenario, a case where energy in particles exceeds the energy in the magnetic
field by a factor of 100, would still require a distance of about 40~kpc for this object
(for $B = 1\, \mu$G).

A situation where the energy density in particles roughly equals the energy
density in the magnetic field for \HESSJ\ would place this object at an extragalactic
distance ($\gtrsim 1$~Mpc). Such a scenario face the challenge to explain the
connected rather large extension of the source. For the case where \HESSJ\ belongs to a 
new class of gamma-ray emitters, evaluation of the distribution of related
objects (if existing) may further help to constrain the environment of HESS~J1507-622.
 
\begin{acknowledgements}
The author acknowledges support from his host institution.
The author wants to thank S. Ohm, P. Eger and A. Franzen 
for stimulating discussions.  
\end{acknowledgements}

\end{document}